\documentclass[superscriptaddress,twocolumn,showpacs,preprintnumbers,amsmath,amssymb,prl]{revtex4}
\usepackage[all]{xy}
\usepackage{graphicx,psfrag,times,epsfig,color}
\usepackage{verbatim}

\begin{document}

\title{Full electrostatic control over polarized currents through spin-orbital Kondo effect}
\author{C.~A. B\"usser} 
\affiliation{Dept. of Physics and Astronomy, University of Wyoming, Laramie, WY 82071, USA.}
\author{A.~E. Feiguin}
\affiliation{Dept. of Physics and Astronomy, University of Wyoming, Laramie, WY 82071, USA.}
\author{G.~B. Martins} 
\email[Corresponding author: ]{martins@oakland.edu}
\affiliation{Department of Physics, Oakland University, Rochester, MI 48309, USA.}

\begin{abstract}
{Numerical calculations indicate that by suitably controlling the individual gate voltages 
of a capacitively coupled parallel double quantum dot, with each quantum dot coupled to one of two 
independent non-magnetic channels, this system can be set into a spin-orbital Kondo 
state by applying a magnetic field. 
This Kondo regime, closely related to the SU(4) Kondo, flips 
spin from one to zero through cotunneling processes that generate almost totally 
spin-polarized currents with opposite spin orientation along the two channels. 
Moreover, by appropriately changing the gate voltages of both quantum dots, one can simultaneously 
flip the spin polarization of the currents in each channel. As a similar 
zero magnetic field Kondo effect has been recently 
observed by Y. Okazaki {\it et al.} [Phys. Rev. B {\bf 84}, (R)161305 (2011)], we 
analyze a range of magnetic field values where this polarization effect seems robust, suggesting that  the 
setup may be used as an efficient bipolar spin filter, which can generate electrostatically 
reversible spatially separated spin currents with opposite polarizations. 
}
\end{abstract}
\maketitle

{\it Introduction.}\textemdash Traditional spintronic devices rely on the use of 
ferromagnetic source and drain leads to produce and detect polarized spin-currents, 
like, for example, the Datta-Das spin field-effect-transistor \cite{sfet}. 
More recently, the manipulation of single spins has become one of the paradigms for 
quantum information. To achieve easier integration with current technology, 
the use of semiconducting lateral single quantum dots (QD) has been suggested as 
a means to produce spin filtering and spin memory devices \cite{recher-loss}, 
which can be controlled through the use of electrostatic gates, 
without the need of ferromagnetic contacts, nor highly inhomogeneous static magnetic fields, or AC fields. 
Its experimental realization \cite{hanson-kouwenhoven}, using a single QD 
and a large magnetic field to produce a {\it bipolar} electrically tunable spin filter, 
has spurred a multitude of proposals, e.g., two QDs embedded in an Aharonov-Bohm ring 
\cite{AB-ring}, a double QD (DQD) in parallel \cite{dahlhaus}, or in a T-shape geometry \cite{mireles-ulloa}, to cite a few.
More related to the results presented here, Borda~{\it et al.} \cite{borda} suggested the possibility of 
spin-filtering in a DQD device at quarter-filling, by exploiting spin and 
orbital degrees of freedom simultaneously through an SU(4) Kondo state. Right after that, Feinberg and Simon \cite{feinberg}, by extending 
the ideas described in \cite{borda} to a similar DQD device, suggested the interesting possibility of a 
``Stern-Gerlach'' spin filter effect at {\it half-filling}. In this work we use two fully independent channels, 
and present detailed numerical results confirming the high efficiency of the spin filtering effect and suggest experimental 
ways of observing it. 

The utilization of the Kondo effect \cite{hewson} in a single-QD \cite{david-kondo} 
has the potential to add an extra dimension to spintronics, as now the localized moment in a QD participates in a 
many-body state that may provide new functionalities to spintronic devices \cite{lindelof}. 
More complex Kondo-like regimes, like the so-called SU(4) Kondo state \cite{loganJPCM,martins1}, 
may provide even additional latitude to create, manipulate, and explore spintronic devices using QDs. 
In this work, we extend a recently observed variant 
of the SU(4) Kondo effect \cite{Okazaki} 
(dubbed the spin-orbital Kondo effect) to propose a device based on a capacitively coupled parallel DQD 
which, when in the Kondo regime (through the application of a magnetic field --- see below), 
functions as a {\it bipolar} spin filter that can produce currents with opposite 
polarities {\it simultaneously} (one in each channel of the DQD system). 
In addition, their polarities can be reversed by tuning the gate voltages of the QDs, 
i.e., the proposed bipolar spin filter is {\it electrically} tunable. As mentioned above, 
a similar device had been suggested before \cite{feinberg}. Here, we provide extensive 
numerical results to stimulate experimental groups to try and observe this effect. 

\begin{figure}[h]
\centerline{\psfig{file=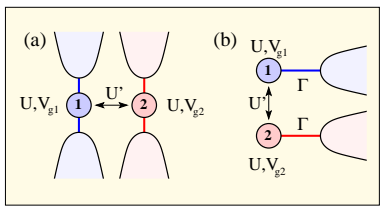,width=5.0cm}}
\caption{(Color online) (a) DQD connected to
four metallic leads so that conductance through them can be measured {\it independently}.
The QDs are subjected to an inter-(intra-)QD Coulomb repulsion $U^{\prime}$ ($U$).
(b) After an even-odd transformation, two of the leads decouple
and the system is reduced to just two leads coupled {\it only} through $U^{\prime}$.
}
\label{figure1}
\end{figure}

{\it Device and Hamiltonian.}\textemdash
The proposed setup is that of capacitively coupled parallel DQDs \cite{Okazaki}  
connected to {\it completely} independent metallic leads [see Fig.~\ref{figure1}(a)] \cite{Amasha}. 
Through an even-odd transformation, two leads decouple from the DQD 
and the system is reduced to that shown in Fig.~\ref{figure1}(b). 
Note that this transformation does not involve the QDs, therefore the interacting part of
the Hamiltonian [(Eq.~(2) below] remains unchanged. 
Then, the two-impurity Anderson Hamiltonian modeling our system is 

\begin{eqnarray}
 H_{\rm tot} &=& H_{\rm DQD} + H_{\rm band}
 + H_{\rm hyb} ,\\
  \label{eq1}
H_{\rm DQD}&=&\sum_{\lambda=1,2;\sigma}
\left[ {U \over 2} n_{\lambda \sigma} n_{\lambda \bar{\sigma}} + (V_{g\lambda}-\sigma H) n_{\lambda \sigma}\right] +  \nonumber \\
  && U^{\prime} \sum_{\sigma \sigma^{\prime}} n_{1 \sigma} n_{2 \sigma'} ,\\
  \label{eq2}
 H_{\rm band} &=&  t \sum_{\lambda=1,2}
 \sum_{i=1;\sigma}^\infty
  (c_{\lambda i\sigma}^{\dagger} c_{\lambda i+1\sigma} +\mbox{H.c.}) , \\
  \label{eq3}
H_{\rm hyb} &=& \sum_{\sigma;\lambda=1,2} t_{\lambda}
\left[ d_{\lambda \sigma}^{\dagger} c_{\lambda1\sigma} + \mbox{H.c.} \right] .
  \label{eq4}
\end{eqnarray}

The operator $d_{\lambda \sigma}^\dagger$ ($d_{\lambda \sigma}$) creates (destroys) 
an electron in QD $\lambda=1,2$ with spin $\sigma=\pm$, while operator $c_{\lambda i \sigma}^\dagger$ 
($c_{\lambda i+1 \sigma}$) does the same at site $i$ ($i+1$) in a non-interacting semi-infinite chain $\lambda=1,2$; 
$n_{\lambda \sigma}= d_{\lambda \sigma}^\dagger d_{\lambda \sigma}$ is the 
charge per spin at each QD, and both QDs have the same charging energy $U$.
We include the effect of a magnetic field $H$ acting just on the QDs \cite{recher-loss}. 
For simplicity, we take the hybridization parameters $t_1=t_2=t^{\prime}$.
It is important to note that, contrary to Ref.~\cite{feinberg}, the {\it only} interaction between electrons in different 
channels $\lambda=1,2$ is the inter-dot capacitive coupling $U^{\prime}$. 
Finally, as our setup consists of semiconducting lateral QDs, each of them can have different gate 
potentials $V_{g1}$ and $V_{g2}$ \cite{Okazaki,Amasha}, and we will concentrate on the experimentally relevant regime $U^{\prime}/U < 1.0$ 
(the so-called $SU(2) \otimes SU(2)$ regime) \cite{loganJPCM}. All results shown were calculated using $U$ as our unit of energy. 
The width of the one-body resonance for each QD is given by $\Gamma= \pi t^{\prime 2} \rho_{0}(E_F)$, 
where $\rho_{0}(E_F)$ is the density of states of the leads at the Fermi energy $E_F$. Throughout the paper $t=1.0$ 
and all the other parameter values are indicated in the figures or in the text.

This model has been studied extensively in previous works,
and it is well known that for $U^{\prime}/U=1.0$ and zero-field it has an SU(4) Kondo fixed point \cite{loganJPCM},
experimentally observed in Refs.~\cite{sasaki,herrero}. Here, we want to address
a completely different regime, although we will also show that
our Density Matrix Renormalization Group (DMRG) \cite{white} calculations faithfully describe the SU(4) Kondo regime as well.

\begin{figure}
\centerline{\psfig{file=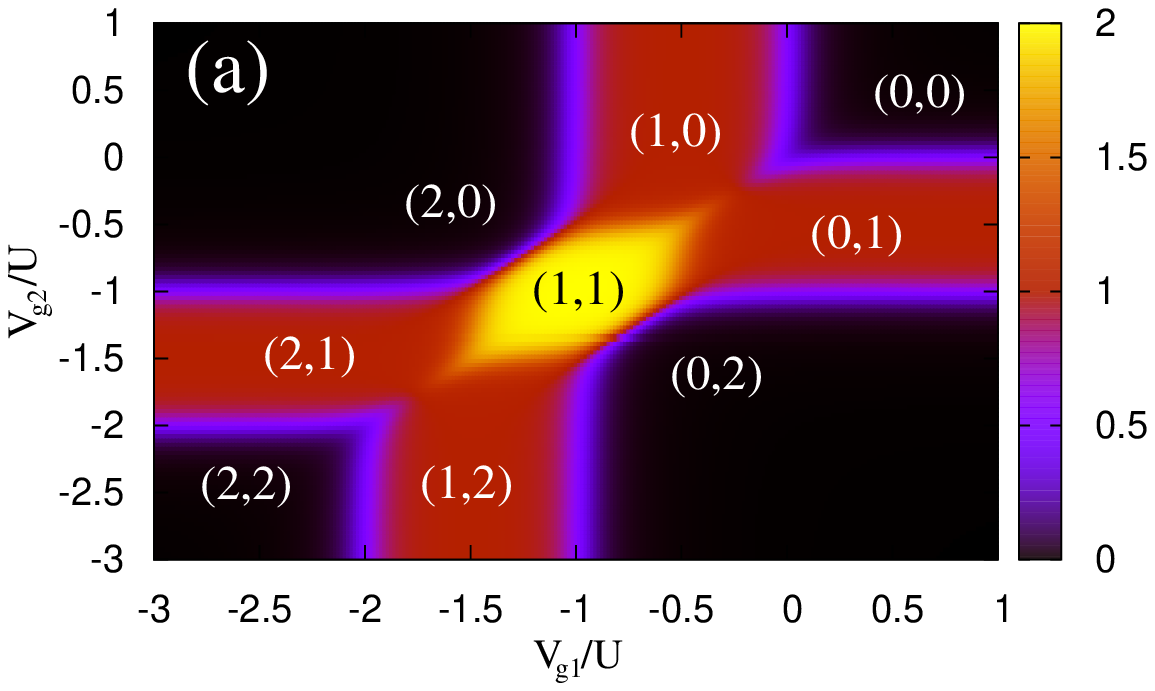,width=7.0cm}}

\vspace{-1.0cm}
\centerline{\psfig{file=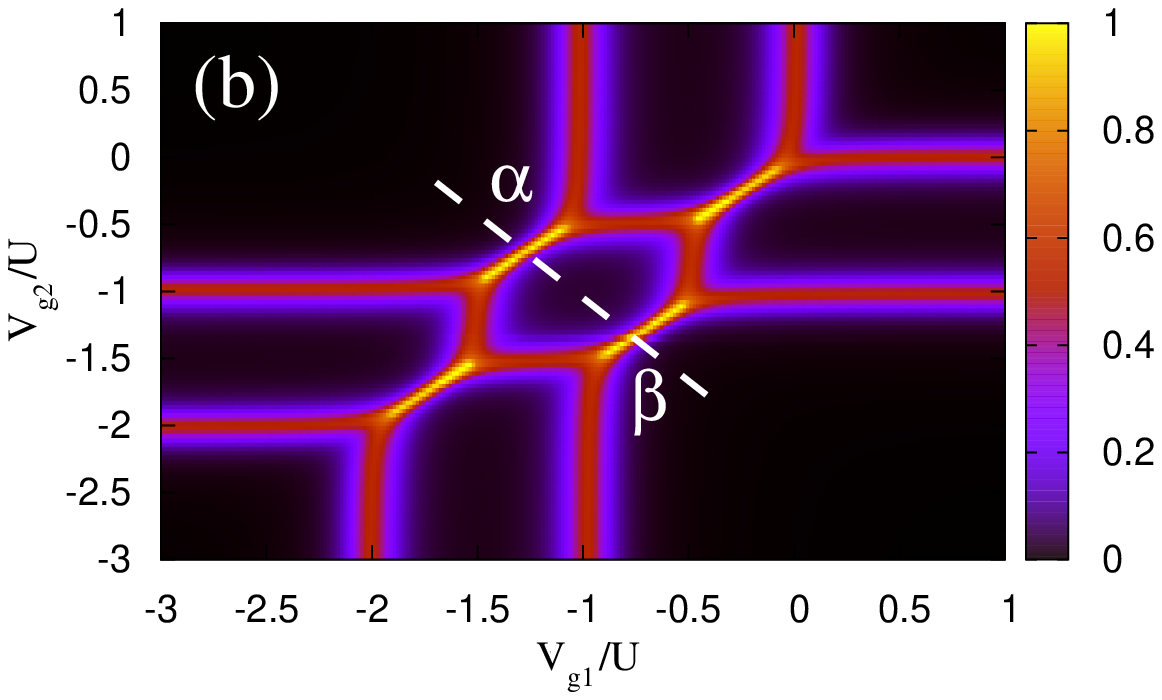,width=7.0cm}}
\caption{(Color online) Conductance (in units of $G_0$) for $U^{\prime}=0.5$, $\Gamma=0.02$, $H=0.0$ (a) and $H=0.05$ (b). 
$\left(n_1,n_2\right)$ specifies the occupancies of each QD
[same values apply to (b)]. (b) $H=0.05$ suppresses the spin Kondo effect
(note that color scales in each panel are different). We show that the bright (yellow) lines
intercepted by the (white) dashed line, where $G=G_0$, 
correspond to a peculiar Kondo effect. Points $\alpha$ and $\beta$ are discussed
in detail in Figs.~\ref{figure3} and \ref{figure4}.
}
\label{figure2}
\end{figure}

{\it Density Matrix Elements.}\textemdash The results presented in this work were calculated using the 
DMRG \cite{dmrg} and the Friedel Sum Rule (FSR) \cite{hewson,Langreth,Logan3,Aligia_FSR}. 
The validity of the FSR for the system studied here is
discussed in the supplemental material \cite{supplem}. 
In order to characterize and identify different regimes, we use the reduced density matrix 
elements (DME), calculated with the DMRG. The ground state wave-function can be written as

\begin{equation}
|\Psi_0\rangle = \sum_{\gamma,\delta} \psi_{\gamma,\delta} | \gamma \rangle | \delta \rangle,
\end{equation}
where $\gamma$ stands for the 16 possible DQD configurations
(0-0, $\sigma$-0, 0-$\sigma$, $\sigma$-$\sigma'$, 2-0, 0-2, $\sigma$-2, 2-$\sigma$, and 2-2), while
$\delta$ represents the states associated with the Fermi sea.
Summing over the band states $\delta$ we obtain the weight projection
of the different DQD configurations in the ground state.
\begin{equation}
\rho_{\gamma,\gamma^{\prime}} = \sum_\delta \psi_{\gamma,\delta} \psi^*_{\gamma^{\prime},\delta}.
\end{equation}
As will be shown in Fig.~\ref{figure4}, the diagonal DME can be used as a `proxy order parameter' for the 
typical correlations that characterize a many-body state like,
for example, the Kondo state. This is very
useful in the case of an unusual (or exotic) Kondo effect, where it may not be clear at first
what are the relevant correlations that one should look for (from now on 
we generally refer to the {\it diagonal} matrix elements as DME weight, or simply DME).

\begin{figure}
\centerline{\psfig{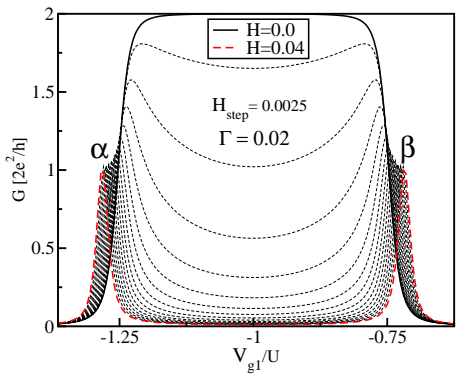}}
\caption{(Color online) Effect of magnetic field over conductance along the (white)
dashed line in Fig.~\ref{figure2}(b). 
$U^{\prime}=0.5$, and $\Gamma=0.02$, $0.0 \leq H \leq 0.04$ (field increases in steps 
of 0.0025). Half-filling conductance (around $V_{g1}\sim -1.0$) in the $SU(2) \otimes SU(2)$ regime 
is suppressed faster than in the SU(4) regime (not shown). The dashed (red) line ($H=0.04$) 
reaches unitary conductance $G=G_0$ at values of $V_{g1}$ corresponding to the points 
$\alpha$ and $\beta$ in Fig.~2(b). 
}
\label{figure3}
\end{figure}

\begin{figure}
\epsfig{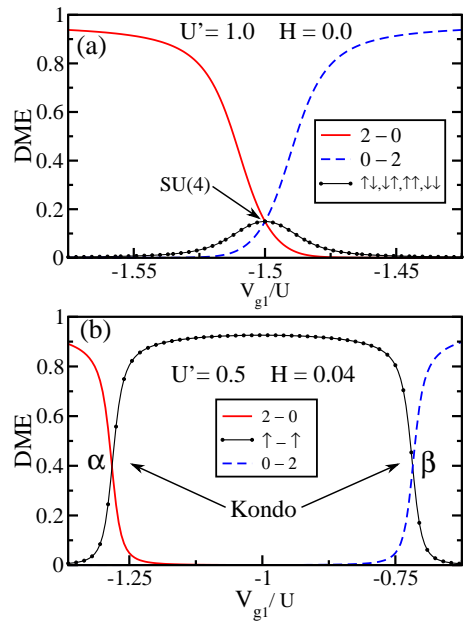}
\caption{(Color online) (a) DME {\it vs.} $V_{g1}$ for $V_{g2}=-V_{g1}-(1+2U^{\prime})$, $\Gamma=0.02$, 
$U^{\prime}=1.0$, and $H=0.0$. At the p-h symmetric point ($V_{g1}/U=-1.5$), we have a half-filling SU(4) Kondo regime, 
characterized by the entanglement of spin and `orbital' degrees of freedom, translated here 
into the equality of all the two-electron DME DQD configurations at $V_{g1}/U=-1.5$. 
(b) Same as in (a), but now for $U^{\prime}=0.5$ ($SU(2) \otimes SU(2)$ regime) and $H=0.04$, 
corresponding to the situation shown for the (red) dashed curve in Fig.~\ref{figure3}. 
Note that the DME of the spin configuration $\uparrow$-$\uparrow$ [(black) dotted curve] is the same as the `orbital' 
configuration $2$-$0$ [(red) solid curve] for $V_{g1}=-1.29$ ($\alpha$ point), as well as for the 
configuration $0$-$2$ [(blue) dashed curve] for $V_{g1}=-0.71$ ($\beta$ point).
}
\label{figure4}
\end{figure}

{\it Numerical Results.}\textemdash Figure \ref{figure2}(a) shows the conductance $G=G_1+G_2$ 
($G_{1,2}$ is the conductance for each channel) 
obtained through the FSR \cite{supplem} in the $V_{g1}-V_{g2}$ plane for $U^{\prime}=0.5$ at zero magnetic field. 
The different QD occupancies 
are indicated by the notation $(n_1,n_2)$. In Fig.~\ref{figure2}(b) we present the conductance results 
for finite field $H=0.05$, where the suppression of spin $SU(2)$ Kondo in each channel can be clearly observed 
[color scales are not the same for panels (a) and (b)]. 
The (white) dashed line is the region of gate voltage variation in 
the $V_{g1}-V_{g2}$ plane that interests us. It is parameterized 
by the expression $V_{g2}=-V_{g1}-(1+2U^{\prime})$. Conductance results along this line 
for $0.0 \leq H \leq 0.04$ are shown in Fig.~\ref{figure3}, where the (black) solid line shows 
results at zero magnetic field, with a well defined plateau around $V_{g1}=-1.0$ 
[it corresponds to a cross section of the bright (yellow) region in Fig.~2(a)]. As the field increases, 
in steps of $\Delta H=0.025$ (dotted lines), the conductance at (and around) the particle-hole (p-h) symmetric 
point ($V_{g1}=-1.0$) is suppressed very quickly, while narrow peaks start to form close to 
the charge degeneracy points [(red) dashed line], denoted $\alpha$ [(2,0)-(1,1)] and $\beta$ [(0,2)-(1,1)] points 
in Fig.~\ref{figure2}(b), where $G = G_0$. Note that these peaks are 
narrow along the (white) dashed line in Fig.~\ref{figure2}(b), but along the charge degeneracy 
lines [the diagonal (yellow) bright lines in Fig.~\ref{figure2}(b)] they present a clear plateau 
structure.

Now, in Fig.~\ref{figure4} we present one of the central results in 
this work. As an illustration of the use of DMEs to trace the possible 
existence of a Kondo regime, panel (a) shows the DME for half-filling ($n_1 + n_2 = 2$) configurations of the DQD 
for $U^{\prime}=1.0$ and zero magnetic field (SU(4) fixed point), 
as a function of $V_{g1}$ for $V_{g2}=-V_{g1}-(1+2U^{\prime})$ 
[equivalent to the dashed (white) line in Fig.~\ref{figure2}(b), but for $U^{\prime}=1.0$ (see supplemental 
material) \cite{supplem}]. At the p-h symmetric point ($V_{g1}=V_{g2}=-U^{\prime}-U/2=-1.5$) 
in Fig.~\ref{figure4}(a), i.e., at the 
half-filled SU(4) fixed point, one sees that the six possible two-electron configurations 
have all the same DME weight in the ground state, 
highlighting the fact that orbital and spin degrees of freedom are perfectly equivalent in the 
half-filled SU(4) Kondo state, i.e., spin 
and orbital degrees of freedom are {\it maximally} entangled \cite{herrero}. This result is well 
known, but it serves to illustrate the use of the DME calculation to `look for' 
possible Kondo states. This is what is done in panel (b), where we present 
the DME results for $U^{\prime}=0.5$ and finite field $H=0.04$ 
(same parameters as the ones for the (red) dashed line in Fig.~\ref{figure3}). In this case, we have two 
different values of $V_{g1}$ for which we have two half-filling configurations 
with the same DME weight (same $V_{g1}$ values as the $\alpha$ and $\beta$ points in Fig.~\ref{figure3}). 
The crossing in the $\alpha$ ($\beta$) point in panel (b) is between configurations $\uparrow$-$\uparrow$ 
and 2-0 (0-2). The important fact to note is that {\it exactly} at these crossings $G=G_0$ 
(see Fig.~\ref{figure3}), indicating the possibility of a Kondo effect. 

\begin{figure}
\centerline{\psfig{file=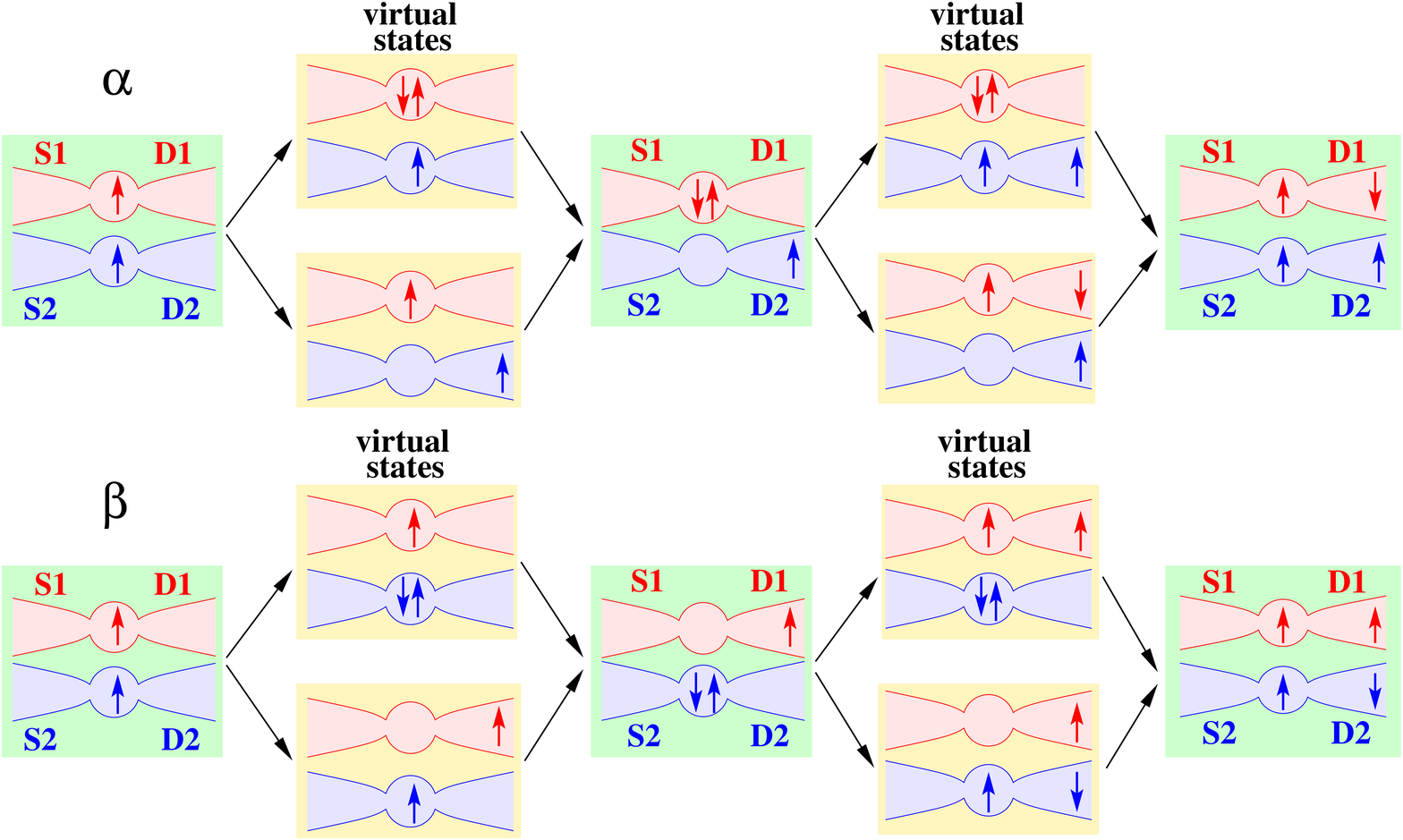,width=08cm}}
\caption{(Color online) Top: Schematic representation of the four possible processes that flip the total spin 
from $S_z=1$ to $S_z=0$, and back to $S_z=1$, when the singlet state is in channel 1 ($\alpha$ point). The net effect is the 
transport of a spin down (up) in channel 1 (2). The coherent superposition of these processes 
leads to Kondo-screening of the pseudospin associated to the two degenerate states with $S_z=1$ and $S_z=0$, 
and the consequent generation of a spin down (up) current in channel 1 (2). Bottom: Equivalent processes for 
the configuration where the singlet state is in channel 2 ($\beta$ point). In this case, 
the polarity of the currents in channels 1 and 2 (when compared to top panel) is reversed. 
}
\label{figure5}
\end{figure}

Indeed, in Fig.~\ref{figure5} we show eight cotunneling processes (four in 
the upper panel and four in the lower) that shift the total $S_z$ spin of the DQD from 
$S_z=1$ to $S_z=0$ (and vice-versa). 
The top processes correspond to the degenerate states $\uparrow$-$\uparrow$ and 2-0 
[$\alpha$ point in Fig.~\ref{figure4}(b)],
while the bottom processes correspond to the degenerate states $\uparrow$-$\uparrow$ and
0-2 ($\beta$ point). The virtual states contain either one or three electrons.
The remarkable fact about these cotunneling processes is that they generate spin polarized
currents in each channel, with opposite polarizations. In addition, once one
sweeps $V_{g1}$ from $\alpha$ to $\beta$, the polarization
direction of the {\it spin filtered} current in each right-side lead is reversed: $\downarrow$ ($\uparrow$) and $\uparrow$ ($\downarrow$) 
in channels 1 and 2, respectively, for the upper (lower) processes. Note that no other 
virtual states are connected (by $t^{\prime}$) to any of the degenerate states 
[$\uparrow$-$\uparrow$ and 2-0 (0-2)] in the $\alpha$ ($\beta$) point \cite{note0}. 
There are similarities between the Kondo effect described here and the one
in Ref.~\cite{Okazaki}: from the lower inset on their Fig.~2 we see that the magnetic field
raises (lowers) the energy of the configuration $\downarrow$-$\downarrow$ ($\uparrow$-$\uparrow$),
while maintaining the configurations $\uparrow\downarrow$-$0$, $\uparrow$-$\downarrow$, and $\downarrow$-$\uparrow$ degenerate 
(to zero-order in $t^{\prime}$).
By adjusting the gate potentials $V_{g1}$ and $V_{g2}$, the configurations $\uparrow\downarrow$-$0$ and $\uparrow$-$\uparrow$
can be made degenerate ($\alpha$ point). Then, the cotunneling processes in Fig.~\ref{figure5} give origin to the Kondo effect
discussed here.

\begin{figure}
\centering
\begin{minipage}[c]{0.5\linewidth}
\centering \includegraphics[width=1.425in]{fig6a.eps}
\end{minipage}%
\begin{minipage}[c]{0.5\linewidth}
\centering \includegraphics[width=1.775in]{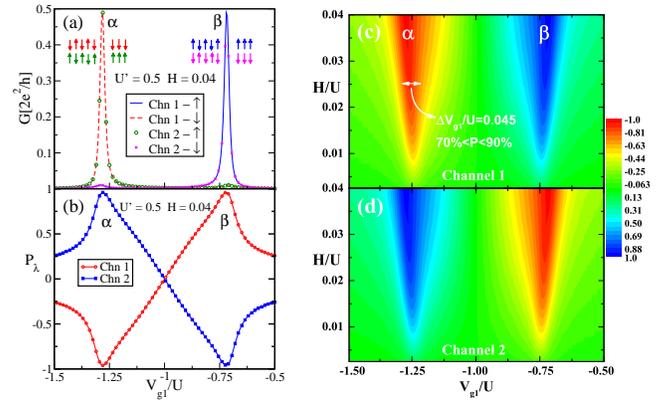}
\end{minipage}
\caption{(Color online) (a) Channel conductance per spin {\it vs} $V_{g1}$ [same interval 
as in Figs.~\ref{figure3} and \ref{figure4}(b), for $H=0.04$ and $\Gamma=0.2$: at point $\alpha$ channel 1 (2) 
is polarized down (up), and at point $\beta$ channel 1 (2) is polarized up (down). 
(b) Polarization $P_{\lambda}$ for each channel calculated using data from (a).
(c), (d) Polarization {\it vs} $V_{g1}$ ($0.0025 \leq H \leq 0.04$) for channels 1 and 2, respectively. 
Panel (c) also shows that for $H/U=0.025$ there is a substantial range $\Delta V_{g1} \approx 0.045$ for which 
the spin polarization varies between $\approx 70\%$ and $\approx 90\%$, indicating that the effect 
should be experimentally observable.
}
\label{figure6}
\end{figure}

In Fig.~\ref{figure6}(a), we show conductance per spin type as a function of $V_{g1}$ 
(same parameters as Fig.~\ref{figure3}, (red) dashed curve) for channels 1 and 2 (see legend). 
These results confirm that the conductance at points $\alpha$ and 
$\beta$ are almost perfectly polarized, in accordance with the cotunneling processes described in Fig.~\ref{figure5}. 
Figure~\ref{figure6}(b) shows the conductance polarization $P_{\lambda}=\left(G_{\lambda \uparrow} - G_{\lambda \downarrow} \right)/
\left(G_{\lambda \uparrow} + G_{\lambda \downarrow} \right)$ for channels $\lambda=1,2$. Panels (c) and (d) show 
polarization results for channels 1 and 2, respectively, for $0.0025 \leq H \leq 0.04$ [other parameters as in panels 
(a) and (b)]. The results clearly indicate that the polarization {\it effect} is robust and does not require a high value of magnetic field. 
Indeed, as indicated in Fig.~\ref{figure5}(c), a polarization of almost $90\%$ can be achieved for $H/U=0.025$. 
Taking $U \approx 1.0meV$ for a GaAs QD\cite{Amasha} results in $H \lesssim 1.0T$ around the $\alpha$ point.  
As indicated by the double-head white arrow in Fig.~\ref{figure5}(c), a range of $\Delta V_{g1} \approx 0.045$, 
at $H/U=0.025$, has a polarization varying from $\approx 70\%$ to $\approx 90\%$. This indicates 
that there is enough range in the parameters space to allow for experimental observation of very high 
polarizations without the need of very high magnetic fields.

{\it Conclusions.}\textemdash In summary, we have presented a peculiar Kondo effect involving a capacitively 
coupled parallel DQD, connected to two independent channels. To achieve this effect it is necessary to 
apply a moderate magnetic field and adjust the gate potential of each QD to take the DQD to a half-filling 
charge degeneracy point. The cotunneling processes in this Kondo effect are such that spin 
polarized currents are generated in each channel, with opposite polarities. The analysis in Fig.~\ref{figure5} 
indicate that the effect should be experimentally observable.

The authors acknowledge very fruitful conversations with Sami Amasha, David Goldhaber-Gordon, Andrew Keller, and Edson Vernek. 
G.B.M. acknowledges financial support by NSF under Grant No. DMR-1107994, and A.E.F. under Grant No. DMR-0955707.

\section{Supplemental Material}

{\it Friedel Sum Rule and Conductance.}\textemdash The first step in calculating 
the conductance through the DQD using the Friedel Sum Rule (FSR) \cite{hewson,Langreth,Logan3} 
is to use the following expression for the conductance \cite{Logan3}
\begin{equation}
G_{\lambda \sigma} = \frac{e^2}{h} \pi \Gamma \rho_{\lambda \sigma}(E_F), 
\end{equation}
where $\Gamma$ is the coupling of each QD ($\lambda=1$, $2$) to its corresponding lead, and $\rho_{\lambda \sigma}(E_F)$ 
is the local density of states (LDOS) of each QD at the Fermi energy. 
The FSR (valid if the ground state of the system is a Fermi liquid, which is true in our case 
as long as $U^{\prime} \leq U$ \cite{loganJPCM}) expresses a relationship between 
$\rho_{\lambda \sigma}(E_F)$ and $n_{\lambda \sigma}$ (QD occupancy), 
\begin{equation}
\rho_{\lambda \sigma}(E_F) = \frac{\cos^2(\pi \left( n_{\lambda \sigma} - 1/2 \right) + I_{\lambda})}{\pi \Gamma},
\end{equation}
where $I_{\lambda}$ is given by
\begin{equation}
I_{\lambda} = \mbox{Im} \int_{-\infty}^{E_F} d\omega~G_{\lambda\lambda}^{\sigma} (\omega) \frac{\partial \Sigma}{\partial \omega},
\end{equation}
and $G_{\lambda \lambda}^\sigma (\omega)$ is the Green's function for each QD, 
\begin{equation}
G_{\lambda \lambda}^\sigma (\omega) = \frac{1}{\omega -\Sigma_{mb}^\sigma(\omega) - \Sigma(\omega)},
\end{equation}
being $\Sigma_{mb}^\sigma(\omega)$ the many-body self-energy, and $\Sigma(\omega)$ the 
one-body part of the self energy, which can be calculated exactly by taking $U=0$.
It can be shown \cite{Aligia_FSR} that, for the parameters used in this work, the integral in the calculation of $I_{\lambda}$ is approximately zero 
(within machine precision).
Therefore, the conductance per channel, per spin, can be calculated as,
\begin{equation}
G_{\lambda \sigma} = \frac{e^2}{h} \sin^2(\pi n_{\lambda \sigma}), 
\end{equation}
and the total conductance, through both channels, is given by
\begin{equation}
G=\sum_{\lambda \sigma} G_{\lambda \sigma}.
\end{equation}
As an example, we will use eqs.~(11) and (12) to calculate the conductance in 
the SU(4) regime (i.e., $U^{\prime}=U$), by using DMRG to calculate the ground state charge per spin $n_{\lambda \sigma}$ 
for each QD for a wide range of $V_{g1}/U$ values. 
We also apply a magnetic field that, contrary to previous works \cite{martins1}, couples 
to the spins in the DQD, but not to the orbital degree of freedom [see eq.~(2)]. 
As there is complete symmetry between orbital and spin degrees of freedom in the SU(4) regime, we should obtain similar results 
to the ones obtained in Ref.~\cite{martins1}.

\begin{figure}
\centerline{\psfig{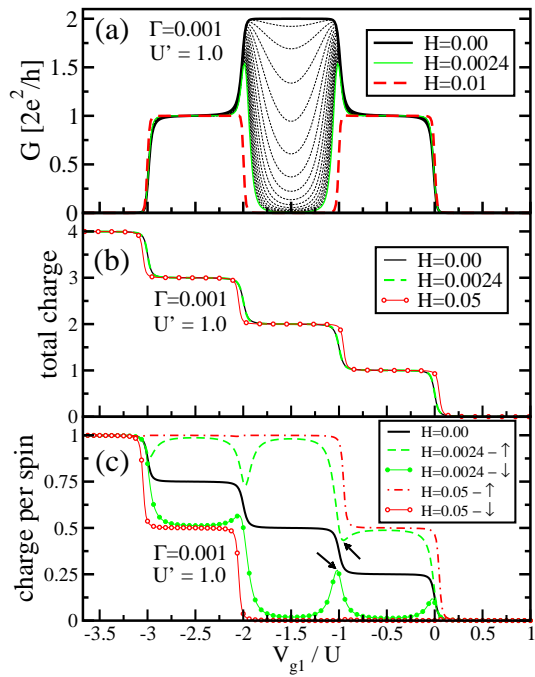}}
\caption{(Color online) (a) Conductance $G$ (in units of $2e^2/h$) and (b) 
DQD total charge as a function of gate potential for $U^{\prime}=U$ (SU(4) regime), 
$V_{g1}=V_{g2}$, and several different values of magnetic
field $H$. In (a), the (black) thick line shows the conductance for $H=0$, while 
thinner (black) dotted lines show the conductance for different $H$ values in steps of $H_{\mbox{\small step  }}=0.0002$.
The thin (green) solid line shows the conductance for $H=0.0024$, while the (red) thick dashed line corresponds to $H=0.05$.
For this large field value, the SU(4) Kondo effect at half-filling is totally suppressed
and just the quarter-filled Kondo effect survives [total charge $1$ and $3$, see panel (b)].
These conductance calculations were done at $T=0$ through the Friedel Sum Rule [see eqs.~(7) to (12)], by calculating the
QD charges, as a function of gate potential, using the DMRG method.
Note that a small $\Gamma=0.001$ value was used to have the system well into the Kondo regime, thus resulting in
very well defined charge plateaus and consequently very well defined conductance plateaus.
Note that in panel (b) well defined plateaus are almost independent of the field $H$,
although, as expected, in panel (c) large variations with field are seen, when the charge {\it per spin}
is plotted as a function of gate potential. It is interesting to
note that for $H=0.0024$ the charge per spin for $V_{g1}/U=0.0$, $-1.0$, and $-2.0$ are similar
(dips and peaks are indicated by arrows for $V_g/U=-1.0$).
This occurs because a small magnetic field is not enough to completely suppress the charge fluctuation (or mixed valence) effects.
We noted that for a larger value of $\Gamma$ the dips and peaks are even more pronounced, as expected (not shown), 
and for a much larger filed ($H=0.05$, (red) dot-dash and open circle curves for spin up and down, respectively), they 
are completely suppressed.}
\label{figure7}
\end{figure}

\begin{figure}
\epsfig{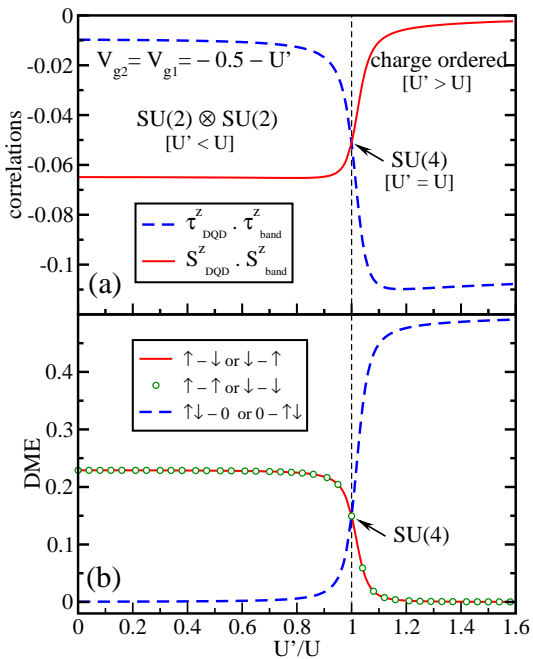}
\caption{(Color online) Spin [(red) solid line] and pseudo-spin [(blue) dashed line] correlations 
(see Ref.~\cite{loganJPCM} for definitions) compared with DME results as a function of $U^{\prime}/U$, 
in panels (a) and (b), respectively. The parameters used were $\Gamma=0.02$ and $V_{g1}=V_{g2}= -0.5-U^{\prime}$ 
(i.e., p-h symmetric point).
In panel (a), the SU(4) point ($U^{\prime}/U=1.0$) is where the curves for the correlations cross, 
while the corresponding point in panel (b) is where the DME for all the half-filling configurations are equal, reflecting the
fact that spin and pseudo-spin degrees of freedom are perfectly entangled. Note also that
the SU(4) point separates different regions: at left, one has an $SU(2) \otimes SU(2)$ ground state,
defined mainly by {\it spin} correlations, while at the right side one has a charge-ordered state, 
defined mainly by {\it charge} correlations \cite{loganJPCM}.}
\label{figure8}
\end{figure}

Indeed, this is what is shown in Fig.~\ref{figure7}(a), where, at zero-field [(black) thick solid line], one sees the three 
characteristic SU(4) conductance plateaus at quarter- and half-filling. For these 
calculations, we have used a small value of $\Gamma=0.001$ to place the DQD deep into the Kondo 
regime (at quarter- and half-filling). As shown previously \cite{martins1}, the magnetic field strongly suppresses 
the Kondo effect at half-filling, but not at quarter-filling (see (black) dotted thin lines), in agreement with the 
fact that orbital and spin degrees of freedom are equivalent in the SU(4) Kondo. The reason being that 
at quarter-filling one is left with an SU(2) {\it orbital} Kondo (in this case, where the magnetic field couples to spins, while 
one was left with an SU(2) {\it spin} Kondo in Ref.~\cite{martins1}, where the field couples to the orbital 
degree of freedom). For $H=0.01$ the SU(4) Kondo has been 
completely suppressed at half-filling [(red) dashed line]. Note that in panel (b) one 
sees very little field dependence in the {\it total}-charge 
variation with gate voltage for the whole range of field explored 
($0.0 \leq H \leq 0.05$). This is deceiving, though, as there is 
obviously considerable dependence with field in the gate voltage charge variation {\it per spin}, 
as shown in panel (c). In it, we see the interesting 
behavior at intermediate values of field ($H=0.0024$, (green) dashed curve for spin up, and (green) solid dots curve for spin down), 
where, for gate voltage values for which the DQD passes through fluctuating valence regimes (i.e., close to charge degeneracy 
points), there are peaks (dips) in the spin down (up) occupancies (see arrows). Calculations (not shown) for larger values of 
$\Gamma$ lead to more pronounced peaks and dips. Finally, note that for higher values of field 
($H=0.05$, for example), the peaks and dips have been suppressed [(red) dash-dot (open circles) line for spin up (down) in panel c]. 

{\it Identifying regimes using DMRG Density Matrix Elements.}\textemdash As a way of demonstrating the use of the Density Matrix 
Elements (DME) to characterize different regimes in the DQD, we will compare DME results with those for specific correlations  
that are usually used for that same purpose. It is known that at half-filling the DQD passes 
through three different fixed points as the ratio $U^{\prime}/U$ varies from zero to slightly above one: SU(2) $\otimes$ SU(2), 
for $0.0 \leq U^{\prime}/U < 1.0$, SU(4), for $U^{\prime}/U=1.0$, and a charge ordered state for $U^{\prime}/U \gtrsim 1.0$ \cite{loganJPCM}. 
Figure \ref{figure8}(a) shows the evolution of spin [(red) dashed line] and pseudo-spin 
[(blue) solid line] correlations as $U^{\prime}/U$ varies from $0.0$ to 
$1.6$ (see Ref.~\cite{loganJPCM} for definitions). 
First, concentrating on the SU(4) point, one sees that both correlations are identical at that point, which is 
indicated by a dashed vertical line that separates the two regions associated to the other two fixed points. For small 
values of $U^{\prime}/U$ the spin correlations dominate, the pseudo-spin correlations being very small. The opposite occurs for 
the $U^{\prime}/U>1.0$ region, indicating the charge ordered state \cite{loganJPCM}. In Fig.~\ref{figure8}(b), we show the corresponding DME results 
for all the half-filling configurations. The results neatly agree with the overall picture obtained through the correlations 
in panel (a), demonstrating once more the use of DMEs to uncover different phases in strongly correlated electron models.


\begin{thebibliography}{99}
\bibitem{dmrg} The DMRG calculations were done keeping typically between 600 states and 1000 states, for clusters with 
up to 40 sites.
\end{thebibliography}
\end{document}